\documentclass[]{aastex631}

\usepackage{gensymb}
\usepackage{tabularx}
\newcolumntype{Q}{>{\hsize=2.0\hsize}X}
\newcolumntype{s}{>{\hsize=.5\hsize}X}

\shorttitle{The challenge of detecting remote spectroscopic signatures from radionuclides}
\shortauthors{Haqq-Misra et al.}

\begin{document}

\title{The challenge of detecting remote spectroscopic signatures from radionuclides}

\correspondingauthor{Jacob Haqq-Misra}
\email{jacob@bmsis.org}

\author[0000-0003-4346-2611]{Jacob Haqq-Misra}
\affiliation{Blue Marble Space Institute of Science, Seattle, WA, USA}

\author[0000-0002-5060-1993]{Vincent Kofman}
\affiliation{NASA Goddard Space Flight Center, Greenbelt, MD 20771, USA}
\affiliation{American University, Washington DC, USA}

\author[0000-0002-5893-2471]{Ravi K. Kopparapu}
\affiliation{NASA Goddard Space Flight Center, Greenbelt, MD 20771, USA}

\begin{abstract}
The characterization of exoplanetary atmospheres through transit spectra is becoming increasingly feasible, and technology for direct detection remains ongoing. The possibility of detecting spectral features could enable quantitative constraints on atmospheric composition or even serve as a potential biosignature, with the sensitivity of the instrument and observation time as key limiting factors. This paper discusses the possibility that future remote observations could detect the presence of radioactive elements in the atmospheres of exoplanets. Such radionuclides could arise from cosmogenic or geologic sources, as well as from industrial sources, all of which occur on Earth. The detection of radionuclides in an exoplanetary atmosphere could reveal important properties about the planet's geology or space environment, and potentially could serve as a technosignature. However, many radionuclides, including those from industrial sources, attach to aerosol or other particles that cannot be remotely characterized. Limited experimental and theoretical spectral data exist for long-lived radionuclides, but the sensitivity required to detect the spectral features of some known radionuclides would be at least several orders of magnitude greater than required to detect the spectral features of molecular oxygen. Present-day remote spectroscopic observing mission concepts at ultraviolet to mid-infrared wavelengths are not sensitive to discern the presence of radionuclides in exoplanetary atmospheres. Interplanetary fly-by or probe missions may be more likely to provide such data in the future.
\end{abstract}

\section{Introduction} \label{sec:intro}

Radionuclides are radioactive forms of elements in Earth's atmosphere, which can arise from abiotic and industrial sources. The most significant geologic sources of radionuclides are the result of the radioactive decay of uranium and thorium rocks, which release radon gas into the atmosphere that then decays into other products. Cosmogenic radionuclides are formed when high-energy cosmic rays {{from deep space break the nuclei of} atmospheric constituents {{into fragments}, which results in a relatively steady-state abundance of radionuclides like $^{14}$C in Earth's atmosphere. Industrial processes can also lead to the formation and release of radionuclides in the atmosphere, which includes the direct and indirect byproducts of nuclear fission and fusion. Geologic sources of radionuclides should be expected in general on other terrestrial planets, to the extent that heavy radioactive elements should be present in planetary crust layers. {{Cosmogenic sources of radionuclides should likewise be expected on other terrestrial planets with atmospheres due to interactions with the near-uniform cosmic ray flux from galactic sources.} The differences in the abundances and diversity of radionuclides can provide important information about the properties of the planet. Industrial radionuclides on Earth are the result of intentionally-deployed technology; {{the detection of such radionuclides on other planets could be potential technosignature candidates, but these could also indicate previously unknown geological processes.} This paper seeks to explore the extent to which future spectroscopic observations could detect or constrain the presence of abiotic or industrial radionuclides in the atmospheres of terrestrial exoplanets. 

Both abiotic and industrial nuclides have been widely used as tracers in atmospheric science, which significantly improved understanding of processes such as troposphere-stratosphere exchange and other timescales for atmospheric circulation \citep[e.g.,][]{rehfeld1995three,lal2012applications,hirose2012uranium}. Geologic radionuclides have their source at the ground and diffuse upward, while cosmogenic radionuclides have a source at the top of the atmosphere and diffuse downward. Cosmogenic radionuclides tend to be relatively homogenous in distribution, while geologic radionuclides tend to show strong contrasts between land and ocean. Industrial radionuclides are temporal in duration and limited in release to a specific location. These properties make the set of abiotic and industrial radionuclides a valuable tool as tracers: given that the half-lives of the radionuclides are known, the relative abundances of one tracer to another can be used to calculate atmospheric transport timescales. Most radionuclides in solid form tend to attach to aerosol or other particles in the atmosphere, which allows for their subsequent analysis in collected samples of rainwater or dust. Other modeling efforts have focused on understanding the dispersion of radionuclides within the troposphere, specifically motivated by emergency response scenarios \citep[e.g.,][]{leelHossy2018review}; however, this study will focus on long-lived radionuclides (with half-lives greater than one day) that could reach the stratosphere and conceivably be detectable through remote spectroscopy. 

Prior work has already demonstrated the feasibility of detecting some exoplanetary radionuclides. \cite{zhang202113co} reported the detection of the $^{12}$C/$^{13}$C isotopologue ratio on the accreting giant planet TYC 8998-760-1 b orbiting a solar-type star, with observations taken by the Spectrograph for INtegral Field Observations in the Near Infrared (SINFONI) at the Very Large Telescope (VLT) in Chile. \cite{zhang202112co} likewise analyzed archival Keck telescope data to measure the $^{12}$C/$^{13}$C isotopologue ratio of the brown dwarf 2MASS J03552337+1133437, which differs from the previous planetary measurements and suggests that stars and planets may have different formation pathways for these isotopologues. Simulations performed by \citet{morley2019measuring} have shown that JWST could measure the D/H ratios of brown dwarfs and giant planets, ranging from 2$M_{\text{Neptune}}$ to 10$M_{\text{Jupiter}}$, by detecting spectral features of CH$_3$D and HDO (where D is deuterium). Other simulations performed by \citet{molliere2019detecting} have similarly shown that high-dispersion spectroscopy with a facility like the Extremely Large Telescope (ELT) could observe the D/H ratio (through CH$_3$D and HDO) as well as the $^{12}$C/$^{13}$C isotopologue ratio for nearby rocky and ice giant planets. \citet{serindag2021measuring} used archival spectra of the M-dwarf GJ 1002 as a proxy for an exoplanet observation to demonstrate that titanium isotopes could be measured on super-Jupiter planets at large angular separation from the host star when observed with facilities like the VLT or ELT. \citet{Kofman2021} assessed the detectability of HDO in an O$_2$ dominated, desiccated atmosphere by studying various enhancements of the the isotope due to atmospheric loss processes. 

This paper {{considers} the possibility of detecting other radionuclides in the atmospheres of terrestrial exoplanets. This overview includes all geological, cosmogenic, and industrial radionuclides that have been measured in any abundance in Earth's atmosphere and have a half-life greater than one day. The abundances of most radionuclides in Earth's atmosphere are low, and the approach in this paper does not attempt to show that Earth-like abundances of any particular radionuclide should be expected or detectable in exoplanetary atmospheres. Instead, the approach in this paper is to examine radionuclides that are known to exist in Earth's atmosphere and identify any spectral features based on available laboratory measurements. This paper does not make any claim or speculation about the likelihood that any particular abiotic or industrial radionuclides will persist at levels above those on Earth today, but instead this analysis seeks to {{explore} the extent to which such planets could be detected and characterized. If such planets do exist, then the identification of radionuclides and measurement of isotopologue ratios could provide valuable information about atmospheric exchange processes on exoplanets while also serving as a way to search for potential atmospheric technosignatures.

\section{Spectra of Known Radionuclides} \label{sec:radionuclides}

Table \ref{tab:radionuclides} provides a list of radionuclides in Earth's atmosphere with a half-life greater than one day, which represent the set of radionuclides that could enter Earth's stratosphere. The sources for each radionuclide are classified as cosmogenic, geological, or industrial. The table also indicates the phase of each radionuclide in the atmosphere, which can persist as a gas, form into a molecule, or attach to aerosol or particles. 

The aim of this section is not to constrain in detail the spectroscopy of the species mentioned above, but to provide an overview of which species might be potential candidates for further investigations, what information is available, and most of all provide a general understanding of the strategies of detecting radioactive species. 

The best way to search for isotopes would most likely be through rotational-vibrational spectroscopy, in which an shift in the energy of the transitions is observed when a different mass is substituted.  In many cases, the spectroscopy of the radioactive nucleotides can be relatively well constrained using theoretical methods, particularly for the bi-molecular species \citep{coxon_improved_2015}. In the case of the energies of rotational or rotational-vibrational transitions, the effect of addition of a neutron (or several) to the nucleus of the atom can be relatively easily calculated using the harmonic oscillator approximation. The same methods have been successfully applied to constrain the ro-vibrational transitions of the several isotopes of CO \citep{li2015rovibrational} or CO$_2$ \citep{huang__reliable_2014}.  Although pure rotational spectroscopy can be used to study the abundance of molecules, neither the star nor the planet is bright enough in this region to provide sufficient background flux to constrain molecular abundances in planetary atmospheres. Instead, this is mostly applied in less dense but more expanded regions of the universe (\textit{i.e.}, molecular clouds \citep{sano_alma_2021} or cometary comae \citep{bockelee-morvan_deuterated_1998}) rather than than in atmospheres. 

Aside from ro-vibrational spectroscopy, atomic spectroscopy provides a method to identify isotopes. Deuterium has a measurable shift in the electronic spectrum of the (re)ionization of the hydrogen atom. Tritium {{(T)} similarly provides a shift  in the energy levels of the electronic levels (H:121.67 nm; D:121.34 nm; T:121.23 nm) \citep{garcia1965energy}. Similarly, a number of atomic transitions of 3-He enable the direct detection of the lighter isotope as well \citep[e.g.,][]{schneider2018nlte}. For heavier atoms, shifts in the electronic transitions will be considerably less since the relative mass difference is much lower. 
Direct emission of electromagnetic radiation from radioactive decay the radioisotopes would be very diagnostic as well, but this would only be possible \textit{in situ}. As such, the rest of this section will focus on ro-vibrational transitions, which is the range that is typically targeted for the study of exoplanet atmospheres. 

Many of the species in the table are not generally gas forming elements which makes these hard to detect, and often understudied. {{At this point, it is useful to consider the preferred host phase for the different elements, which will guide the likelihood of finding it in the gas phase versus bound in the solid phase. Here, we follow the Goldschmidt classification, often used in geochemistry \citep{goldschmidt1937principles}. Elements known as \textit{Lithophiles} are those that} react readily with oxygen {{to form} ``rock-like'' minerals. {{Lithophile elements tend to be enriched in Earth's crust compared to solar system abundances. Elements known as \textit{Siderophiles}} have strong affinity for iron, {{which} form metallic bonds with iron or other iron-like species {{that} similarly prevents them from being detectable in the gas phase. {{Siderophile elements have a much higher abundance in Earth's core compared to the crust.} Both types of elements are most likely to end up in dust-like particles, from which the composition is nearly impossible to discern by remote observations. When elements bind directly with hydrogen or oxygen, they may be more likely to be able to become volatile enough to be in a gaseous form. The next section discusses the more relevant species in detail.

\begin{table}
    \centering
    \scriptsize
    \begin{tabularx}{0.8\textwidth}{ssXQ}
        \hline
        Radionuclide & Half-life & Phase in Atmosphere  & Relevance   \\
        \hline\hline
        \multicolumn{4}{l}{Cosmogenic} \\
        \hline
        $^{3}$H    & 12.3\,yr   &  molecular (e.g., H$_2$O)  &  \textit{see text} \\
        $^{3}$He   & stable     &  gas  &\\
        $^{7}$Be   & 53\,d      &  aerosol/particles  &  $^{9}$Be abundance in crust $\sim$2 ppm and not a major fission product.   \\
        $^{10}$Be  & 1.5\,Myr   &  aerosol/particles  &   \\
        $^{14}$C   & 5.7\,kyr   &  molecular (e.g., CO$_2$)  & \\
        $^{26}$Al  & 0.71\,Myr  &  aerosol/particles  & Important source of heat in early solar system \cite{urey1955cosmic}, lithophile.   \\
        $^{32}$Si  & 150\,yr    &  molecular (e.g., SiO$_2$)   & Lithophile, but can also be found in gas phase. \\
        $^{32}$P   & 14.3\,d    &  aerosol/particles/gas& Relatively short-lived \\
        $^{33}$P   & 25.3\,d    &  aerosol/particles  &  \\
        $^{37}$Ar  & 35\,d      &  gas  & Noble gas, only forms molecules under most extreme conditions   \\
        $^{39}$Ar  & 268\,yr    &  gas  &   \\
        $^{81}$Kr  & 0.23\,Myr  &  gas  & Noble gas, only forms molecules under most extreme conditions  \\
        $^{129}$I  & 15.7\,Myr  &  aerosol/particles  & Forms several gasous compounds (HI, I$_2$, IO, CH$_3$I), also industrial  \\
        \hline
        \multicolumn{4}{l}{Geological} \\
        \hline
        $^{210}$Pb & 22.2\,yr   &  aerosol/particles  & Volatile, found widely throughout the atmosphere when used in fuel. \\
        $^{210}$Bi & 5\,d       &  aerosol/particles  &   \\
        $^{210}$Po & 138\,d     &  aerosol/particles  &   \\
        $^{222}$Rn & 3.8\,d     &  gas  &Noble gas   \\
        $^{224}$Ra & 3.6\,d     &  aerosol/particles  & Alkali metal, very reactive and possibly volatile \\
        $^{226}$Ra & 1.6\,kyr   &  aerosol/particles  &   \\
        $^{228}$Ra & 5.8\,yr    &  aerosol/particles  &   \\
        $^{228}$Th & 1.9\,yr    &  aerosol/particles  & Lithophile, although reactive, is likely to form heavy and refractive compounds (\textit{i.e.}, TeO$_2$ has melting temperature of 3390$^\circ$C)   \\
        $^{230}$Th & 77\,kyr    &  aerosol/particles  &   \\
        $^{232}$Th & 14\,Gyr    &  aerosol/particles  &   \\
        $^{234}$U & 244\,kyr    &  aerosol/particles  & Similar to Th, reactive, but forms heavy refractory compounds.   \\
        $^{235}$U & 700\,Myr    &  aerosol/particles  &   \\
        $^{238}$U & 4.47\,Gyr   &  aerosol/particles  &   \\        
        \hline
        \multicolumn{4}{l}{Industrial}\\
        \hline
        $^{60}$Co  & 5\,yr      &  aerosol/particles  & Siderophile   \\        
        $^{90}$Sr  & 29\,yr     &  aerosol/particles  & Lithophile, naturally found in SrSO$_4$ and CO$_3$   \\
        $^{99}$Tc  & 210\,kyr   &  aerosol/particles  &   \\        
        $^{129}$I  & 15.7\,Myr  &  aerosol/particles  & Forms several gasous compounds (HI, I$_2$, IO, CH$_3$I), also cosmogenic \\
        $^{131}$I  & 8\,d       &  aerosol/particles  &   \\
        $^{133}$Xe & 36\,d      &  gas  &Noble gas  \\
        $^{135}$Xe & 5.3\,d     &  gas  & \\
        $^{137}$Cs & 30\,yr     &  aerosol/particles  & Alkali metal, chemically similar to Na, K. Can form Cs$_2$O or CsH, possibly volatile. \\        
        $^{238}$Pu & 87.7\,yr   &  aerosol/particles  &   \\
        $^{239}$Pu & 24\,kyr    &  aerosol/particles  & Similar to Th, reactive, but forms heavy refractory compounds.   \\
        $^{240}$Pu & 6.6\,kyr   &  aerosol/particles  &   \\
        $^{241}$Am & 432\,yr    &  aerosol/particles  & Synthetic element, although hypothesized to form during natural fission.   \\   
        \hline
    \end{tabularx}
    \caption{Radionuclides in Earth's atmosphere with half-lives greater than 1 day.}    \label{tab:radionuclides}
\end{table}

\subsection{H-3; Tritium}
{{Tritium is one of the more common products from nuclear fission and fusion reactions. It can also be used are as a potential fuel for nuclear fusion \citep{tanabe_introduction_2017}. The atmospheric inventory of T peaked at around 3 orders of magnitude above the background levels in the 1970s to around $5 \times 10^{5}$ g due to nuclear testing, see \citep{happell_history_2004} for an historic overview. Considering the total mass of the atmosphere is on the order of $5 \times 10^{21}$ g, this provides an indication of the detectability challenge. Additionally, the spectroscopy of tritium has not been constrained in great detail, as its natural abundance is so low and its radioactivity provides additional regulatory challenges. In Earth's atmosphere, T is predominantly found in the form of HT and HTO. For HTO, limited broad-band infrared spectroscopy is available from \citet{o2007radiochemical} and \citet{nicodemus2007infrared}. High-resolution studies are ongoing to characterize its full spectroscopy \citep{hermann_HTO,hermann_observation_2023}. Rovibrational transitions of HT have been characterized by \citet{cozijn_precision_2024}.  However, the low abundance of tritium, even at highly `productive' times, make it unlikely to be detectable in remote observations. }
 
\subsection{Carbon-14}
Radioactive carbon is formed in Earth's atmosphere through neutron capture by atomic nitrogen above heights of 9 km in a process called cosmic ray spallation. The equilibrium abundance of $^{14}$C is on the{{ order of 1 ppt \citep{obodovskiy_chapter_2019}}. Because of its long half-life time, it is used in radiocarbon dating, where emission of high-energy electrons is counted as a proxy for the amount of $^{14}$C in the sample (or more recently using mass spectrometry). In addition, the production of $^{14}$C also occurs in nuclear fission, and because of the atmospheric testing of thermonuclear bombs the abundance in the atmosphere roughly doubled until the Partial Nuclear Test Ban Treaty of 1963 and other {{agreements banned such testing \citep{nydal_distribution_1965,caldeira_predicted_1998}.} In terms of the spectroscopy, $^{14}$C substitution would provide a significant shift in the ro-vibrational transitions. Isotopic information for $^{13}$C is available in the HITRAN database, which could be used as a starting point to constrain the spectroscopy of the $^{14}$C isotopes of carbon. Note, however, that the abundance of $^{14}$C is approximately 10 orders of magnitude lower than that of $^{13}$C in Earth's atmosphere, making it unlikely to be detectable. 

\subsection{Silicon-32}
This heavy isotope of silicon is formed, similarly to $^{14}$C,  by cosmic ray spallation. Its spectroscopy may be revealed by including an isotopic shift to any of the Si-containing molecules that are considered candidates to be found in exoplanet atmospheres, such as SiO$_2$ \citep{owens2020exomol}, SiH$_2$ \citep{clark2020high}, SiH$_4$ \citep{owens2017exomol}, SiS \citep{upadhyay2018exomol}, and SiO \citep{barton_exomol_2013}. {{Spectra for a number of these molecules with $^{29}$Si and $^{30}$Si are available: SiS \citep{upadhyay2018exomol}, SiH \citep{yurchenko_exomol_2018}, and SiO \citep{barton_exomol_2013}.}

\subsection{Phosphorus-32/33}
The two radioactive isotopes of phosphorus have relatively short life times and are both artificially produced for medical/scientific applications. The spectrum of $^{31}$PH$_3$ is described in HITRAN \citep{gordon2022hitran2020,muller_cologne_2005}, and several other P containing compounds are available in the ExoMol repository \citep{tennyson2012exomol,prajapat_exomol_2017}. No rovibrational spectra of $^{32}$P or $^{33}$P are available.

\subsection{Iodine-129/131}
Both radioactive isotopes of iodine could potentially be spectroscopically detected in the HI molecule if their abundances were high enough. Chemically, iodine behaves similar to chlorine, with its stable isotopes 35 and 37-Cl being spectroscopically easily distinguishable in a ro-vibrational spectrum (indeed it is a textbook example of isotopic shifts in spectroscopy), and the isotopic shift of the iodine isotopes would be easily calculated for a di-atomic molecule. The spectra of HI and ICH$_3$, both considering the stable isotope $^{127}$I, are available HITRAN \citep{gordon2022hitran2020,goldman_improved_1998,raddaoui_line_2019}. In Figure \ref{fig:detectability} a crude estimation of the detectability is given. For more details see the next section.

\subsection{Lead-210}
Although lead has been widely spread throughout Earth's cryosphere as a form of an industrial pollutant, it is known to bind in several different species, most of these salt-like (\textit{i.e.}, \citet{funasaka_different_2013} list 10 difference compounds). Most likely, these would end up in small particulate matter, preventing remote spectroscopic characterization. 

\subsection{Alkali metals}
The alkali metals highlighted in Table \ref{tab:radionuclides} (K, Rb, and Cs) are all reactive metals that may form several different types of potential gaseous compounds. In terms of the spectroscopy, the spectra specific to the radioactive nuclides could be calculated using quantumchemical methods. The primary limitation, however, is that these compounds have not been well studied, aside from optical (\textit{i.e.}, electronic) spectroscopy.

\section{Detectability Limits}

Attempting to detect molecular spectral signatures in exoplanetary atmospheres remains an ongoing challenge, and extending this idea to the characterization of radionuclides in terrestrial exoplanetary atmospheres will inevitably increase the sensitivity requirements. An illustration of these detectability limits is shown in Figure \ref{fig:detectability}, in which the detectability of HI is assessed. Iodine was one the main contaminants from the {{Chernobyl disaster in terms of radiation \citep{radiation_sources_2011}}, and one of the species which may be the most detectable out of the listed species in Table \ref{tab:radionuclides}. Earth's atmosphere contains around 10 ppt I$_2$ \citep{saiz-lopez_atmospheric_2012}. Iodine {{is found in} several different compounds, with sources rangin{{g from organic \citep{saiz-lopez_novel_2004} to inorganic \citep{schonhardt_observations_2008}}. However,  {{for this assessment we} assume an upper limit on the detectability by assuming HI is the dominant form, since its spectroscopy is accessible, {{and the spectroscopic shift from the isotopic substitution would be much stronger in HI than in IO, for instance.} HI is not the most stable form of iodine and in reality its abundance would be a significantly lower.  For the simulations, we calculate the detectability at 400 ppm HI, which is \textit{at least} seven orders of magnitude higher than realistic, but the radiative transfer model contains a backstop which prevents spectroscopic transitions of too low intensities from being considered. At the 400 ppm level, the absorption still strongly overlaps with other spectroscopic features, so the strength of the absorption is determined without any other species in the atmosphere (see the top panel of Figure \ref{fig:detectability}). The equivalent width is 0.04 $\mu$m, which is a factor of four below CO$_2$ around 4.3 $\mu$m. As such, the estimate is that the detectability is at a minimum eight orders magnitude lower than CO$_2$ in this range.  For comparison purposes, the equivalent width of a number of species is given in Table \ref{tab:Eq_widths}. Note that this {{assessment} only considers the main isotope, not the radionuclide. Considering the fission yields from U$^{235}$ of I$^{129}$ and  I$^{131}$ are 0.7\% and 2.9\% respectively, this would suggest that significant amounts of nuclear material would need to be split even for local enhancements to be detectable. 

\begin{table}
    \centering
    \begin{tabular}{ccc}
            \hline
            Molecule    & wavelength & equivalent width\\
            \hline \hline
            O$_2$       &  0.76     & 0.127\\
            H$_2$O      &  0.85     & 0.827\\
            CH$_4$      &  1.6      & 2.7e-4\\
            CO$_2$      & 4.3       &0.13\\
            HI          & 4.5       &4e-8\\
            \hline
    \end{tabular}
    \caption{Equivalent width of a number of species in the Earth's atmosphere. For this calculation, the US standard Earth's atmosphere was selected. HI was added at an abundance of 400 ppm.}
    \label{tab:Eq_widths}
\end{table}

\begin{figure}[ht!]
\centerline{\includegraphics[width=6.0in]{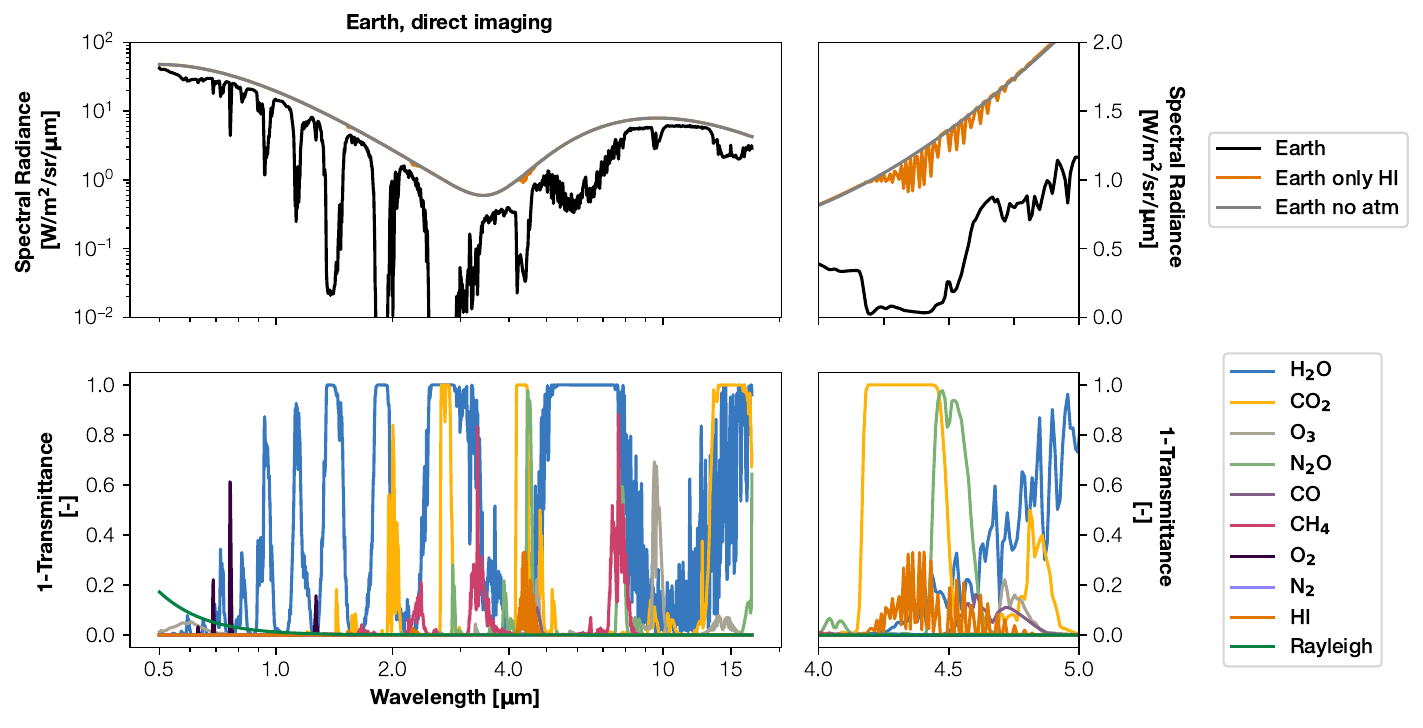}}
\caption{Assessment of the detectability of HI at an enriched abundance (400 ppm versus \textit{at most} 10 ppt). The top panel shows the emitted radiation of the Earth seen at quadrature with the atmosphere (blue), without the atmosphere (gray), and with exclusively 400 ppm of HI in the atmosphere to indicate the positions of the HI spectral features (orange). The bottom panel shows 1-transmittance, which guides the reader through the different absorption features seen in the top panel. HI is indicated in dark orange. The right panels show the same quantities, but focused on the 4 to 5 $\mu$m region which is analysed here. \label{fig:detectability}}
\end{figure}

Detecting radionuclides in exoplanetary atmospheres may be more promising for interstellar fly-by missions \citep[such as Breakthrough Starshot,][]{parkin2018breakthrough} or interstellar probe missions \citep[e.g.,][]{matloff2006deep}, rather than remote detection. Close approaches or orbit around an exoplanet could enable much higher-resolution spectroscopy for atmospheric characterization, while descent probes sent into an exoplanetary atmosphere could provide in situe measurements of a wide range of atmospheric properties (including sample collection and analysis). Such capabilities for remote atmospheric characterization are being developed in the ongoing exploration of the solar system. A long-term strategy for interstellar exploration could be capable of detecting some of the radionuclide spectral features discussed in this paper, but such missions remain in conceptual stages at present. {{It is also worth noting that any exoplanet to be visited by an interstellar flyby or probe mission will likely already have been the subject of extensive remote observations, so the utility of radionuclides as a way to infer atmospheric properties or search for technosignatures may be limited by the time an interstellar mission were to arrive at its destination.}

Remote detection of radionuclide spectra appears challenging, although such conclusions are drawn from limited laboratory and theoretical data. A more complete set of data from either measurements or theory for the radionuclides discussed in this paper would reveal the extent to which any of these species would exhibit uniquely recognizable spectral features that could aid in detection. Conducting such laboratory experiments would be instructive for understanding the extent to which molecular radionuclides can be identified through high-resolution spectroscopy. 

But further experiments might indicate that characterization of radionuclide spectra remains out of reach for even the next generation of space telescopes, which would indicate that any abiotic or industrial radionuclides that might exist in the atmospheres of known exoplanets would be beyond remote characterization with known technology. This paper does not necessarily conclude that such detections will always be impossible, as numerous advances in astronomy (including the mere detection of exoplanets) had historically been regarded as impossible. The purpose of this paper is only to note the detectability limits that would be needed to detect the presence of radionuclides, however difficult such detection may be. Earth itself remains the best analog for an exoplanet, and any attempt at developing the tools for detecting atmospheric radionuclides may perhaps find the greatest success by beginning on Earth and looking up.

\section{Conclusion}

The detection of radionuclides in exoplanetary atmospheres is a formidible challenge, due to the increased sensitivity requirements to distinguish spectral features from isotopes as well as the likely low abundances of atmospheric radionuclides. The overview provided in this paper is based on the limited availability of radionuclide species in spectral databases, so further laboratory measurements and theoretical calculations are needed to better understand the detectable properties of long-lived radionuclides. Nevertheless, many radionuclides may be difficult to detect, particularly because many attach to aerosol or other particles that could not be remotely characterized. Fly-by or probe missions sent to exoplanetary systems may have the greatest opportunity for such high-resolution atmospheric characterization, but future space telescopes operating at ultraviolet to mid-infrared wavelengths may be unlikely to discern such features.

The search for technosignatures would likewise require close fly-by or \textit{in situ} observations to identify the presence of industrial radionuclides in an exoplanetary atmosphere. The presence of such industrial radionuclides in an exoplanet environment would indeed be compelling as a technosignature candidate, but even in the most extreme cases of abundant industrial radionuclides, detecting any spectral features of such extarterrestrial technology will remain beyond the scope of any planned remote observing programs.

\begin{acknowledgments}
J.H.M. and R.K.K. gratefully acknowledge support from the NASA Exobiology program under grant 80NSSC22K1009. The authors also acknowledge support from the Goddard Space Flight Center (GSFC) Sellers Exoplanet Environments Collaboration (SEEC), which is supported by the NASA Planetary Science Division's Research Program. Any opinions, findings, and conclusions or recommendations expressed in this material are those of the authors and do not necessarily reflect the views of their employers or NASA.
\end{acknowledgments}


\bibliography{main}{}
\bibliographystyle{aasjournal}

\end{document}